\documentclass[letterpaper, 10 pt, conference]{ieeeconf}  

\IEEEoverridecommandlockouts                              

\usepackage{amsmath} 
\usepackage{amssymb}  
\usepackage{hyperref}
\usepackage{cite}

\usepackage{textcomp}
\usepackage{stfloats}
\usepackage{url}
\usepackage{verbatim}
\usepackage{graphicx}
\usepackage[linesnumbered,ruled,vlined]{algorithm2e}
\usepackage[caption=false,font=normalsize,labelfont=sf,textfont=sf]{subfig}
\usepackage{pifont}
\usepackage{booktabs}
\usepackage{multirow}
\usepackage{hhline}
\usepackage[table]{xcolor}

\RestyleAlgo{ruled}
\SetKw{Initial}{Initialization:}
\SetKw{Linear}{Linearization:}
\SetKw{History}{Historical Estimates:}
\SetKw{Measurements}{Preparing Measurements:}
\SetKw{System}{Preparing System Statistics:}
\SetKw{Solution}{Matrices for Recursive Solution:}
\SetKwInput{Horizon}{Move Horizon}

\newtheorem{theorem}{Theorem}
\newtheorem{lemma}{Lemma}
\newtheorem{corollary}{Corollary}
\newtheorem{remark}{Remark}

\title{\LARGE \bf Receding Horizon Recursive Location Estimation*
}

\author{Xu Weng$^{1}$, K.V. Ling$^{1}$ and Ling Zhao$^{2}$ 
\thanks{*This work was supported by the NTU Research Scholarship.}
\thanks{$^{1}$Xu Weng and K.V. Ling are with the School of Electrical and Electronic Engineering, Nanyang Technological University, 637798, Singapore
        {\tt\small xu009@e.ntu.edu.sg, ekvling@ntu.edu.sg}}%
\thanks{$^{2}$Ling Zhao is with the School of Mechanical Engineering, Nanjing University of Science and Technology, Nanjing, 210094, China
        {\tt\small zhaoling\_2022@163.com}}%
}

\begin{document}

\maketitle
\thispagestyle{empty}
\pagestyle{empty}

\begin{abstract}
This paper presents a recursive solution to the receding or moving horizon estimation (MHE) problem for nonlinear time-variant systems. We provide the conditions under which the recursive MHE is equivalent to the extended Kalman filter (EKF), regardless of the horizon size. Theoretical and empirical evidence is also provided. Moreover, we clarify the connection between MHE and factor graph optimization (FGO). We apply the recursive MHE to GNSS localization and evaluate its performance using publicly available datasets. 
The paper is based on the deterministic least squares framework.
\end{abstract}

\section{INTRODUCTION}
Localization using sensor measurements is generally a state estimation problem for a nonlinear time-variant (NLTV) system. The classical extended Kalman filter (EKF) is a standard approach for handling such nonlinear estimation tasks \cite{spilker1996global,9101092,thrun2005probabilistic}. Numerous variants of the EKF have been designed to improve its performance \cite{sykacek2002adaptive,rawlings2006particle,gustafsson2011some,lambert2022continuous}. Besides, with the rapid evolution of computing hardware, factor graph optimization (FGO) is also widely applied in various localization systems \cite{dellaert2017factor,wen2021towards,suzuki2023precise}. However, few attentions are drawn to the receding or moving horizon estimation (MHE)\cite{weng2023localization,liu2023state}, another state estimator with a long history \cite{muske1993receding,ling1999receding}. This paper applies MHE to GNSS localization and investigates its connections to the classical EKF and the popular FGO within this context. 


The EKF solves a nonlinear estimation problem by linearizing it at certain approximate states and computing the least squares solution \textit{recursively} \cite{muske1993receding}. It runs quickly but may generate suboptimal solutions due to the inaccuracy of the linearization \cite{rawlings2006particle}.
FGO handles the nonlinearity in an optimization manner --- it linearizes the problem and optimizes the solution \textit {iteratively}. Therefore, despite heavier computational loads, it usually generates better solutions than EKF \cite{wen2021towards}. MHE employs a sliding window of measurements (from a historical time to the current one) to estimate the current system state. This paper indicates that MHE can unify both EKF and FGO under particular prerequisites.

The recursive solution to an MHE problem for a linear time-invariant (LTI) system has been given in previous works. And the equivalence between MHE and Kalman Filter (KF) for LTI systems has been well established since 1993 \cite{muske1993receding,muske1995nonlinear,ling1999receding,rao2001constrained,rawlings2006particle}. However, for NLTV systems, fewer solid studies have explored the unification between MHE and EKF. The existing discussions on this topic stalled in the case of MHE with a zero horizon size \cite{robertson1996moving,rao2002constrained,rao2003constrained}. In this work, we establish the conditions that unify EKF and MHE for any horizon size, demonstrating that EKF is a special case of MHE for NLTV systems.

Factor Graph Optimization (FGO) has been going viral in the GNSS community since its leading performance in the Google Smartphone Decimeter Challenge (GSDC) \cite{ suzuki2023precise}. FGO stems from factorizing a global a posteriori density to perform the maximum a posteriori (MAP) estimation \cite{dellaert2017factor}, which is exactly the same as that done by MHE and will be explained in the following sections.

Our contributions are summarized as follows:
\begin{itemize}
\item We provide the recursive solution to a receding horizon state estimation problem for an NLTV system.
\item We elucidate the connections among EKF, MHE, and FGO.
To the best of our knowledge, our work is the first to provide the conditions under which EKF and MHE achieve equivalence for any horizon size in NLTV systems. 
\item We apply MHE to GNSS localization and evaluate its performance on the public Google Smartphone Decimeter Challenge (GSDC) datasets. 
\end{itemize}
Instead of finding the best estimator, this work focuses on investigating the interconnections among EKF, FGO, and MHE. We provide our perspectives both theoretically and empirically, clarifying the pros and cons of these estimators.

\section{BACKGROUND}
\subsection{System Models}
Consider the following nonlinear time-variant (NLTV) discrete-time system model:
\begin{IEEEeqnarray}{rCl}
\mathbf{x}_{k+1}&=&f_k(\mathbf{x}_k)+\mathbf{w}_k \label{eqn:dynEqn}
\\
\mathbf{y}_k&=&h_k(\mathbf{x}_k)+\mathbf{v}_k\label{eqn:measEqn}
\end{IEEEeqnarray}
where the system state at the $k^{th}$ time step is denoted by $\mathbf{x}_k$. The corresponding observation or measurement about the system is $\mathbf{y}_k$. The nonlinear time-variant function $f_k$ describes the system dynamics, while $h_k$ models the relationship between the state and the measurement. $\mathbf{w}_k$ and $\mathbf{v}_k$ denote state and measurement noise/interference, respectively. We eliminate the control input for simplification.

\subsection{Receding Horizon State Estimation}
The motivation for using a receding horizon for state estimation is to reduce the heavy computational loads when all historical measurements are leveraged over time. The moving horizon estimation (MHE) estimates the current state based on a fixed-size receding/moving horizon/window of historical measurements and can be formulated as a maximum a posteriori (MAP) estimation \cite{muske1993receding,ling1999receding}:
\begin{equation}\label{eqn: MAP_MHE}
\max{p(\mathbf{x}_{k-N},\cdots,\mathbf{x}_k|\mathbf{y}_{k-N},\cdots,\mathbf{y}_k)}.
\end{equation}

Assuming the states $\{\mathbf{x}_{k-N},\cdots,\mathbf{x}_k\}$ are Markov, and the measurements $\{\mathbf{y}_{k-N},\cdots,\mathbf{y}_k\}$ are independent, we can factorize the global a posteriori density in \eqref{eqn: MAP_MHE} as \cite{jazwinski1970stochastic}:
\begin{align}
&p(\mathbf{x}_{k-N},\cdots,\mathbf{x}_k|\mathbf{y}_{k-N},\cdots,\mathbf{y}_k) \nonumber
\\
&\propto p(\mathbf{x}_{k-N})\cdot\prod_{j=k-N}^{k-1} p(\mathbf{x}_{j+1}|\mathbf{x}_j)\cdot\prod_{j=k-N}^{k} p(\mathbf{y}_{j}|\mathbf{x}_{j})\nonumber
\\
&=p(\mathbf{x}_{k-N})\cdot\prod_{j=k-N}^{k-1} p(\mathbf{w}_{j})\cdot\prod_{j=k-N}^{k} p(\mathbf{v}_{j}). \label{eqn:PBLS}
\end{align}

Under the following assumptions of the system statistics:
\begin{gather}
\mathbf{x}_{k-N}\sim N(\hat{\mathbf{x}}_{k-N|k-N-1}, \mathbf{P}_{k-N})\nonumber
\\
\mathbf{w}_j \sim N(0, \mathbf{Q}_{j}), \mathbf{v}_j \sim N(0, \mathbf{R}_{j}),\label{eqn:noiseAssump} 
\end{gather}
where $\mathbf{P}_{k-N}$ is the covariance matrix of the a priori estimation error $\mathbf{w}_{k-N-1}$, $\{\mathbf{Q}_j\}_{k-N}^{k-1}$ and $\{\mathbf{R}_j\}_{k-N}^{k}$ are the covariance matrices of the system noise $\{\mathbf{w}_j\}_{k-N}^{k-1}$ and $\{\mathbf{v}_j\}_{k-N}^{k}$, then MHE can be formulated as:
\begin{IEEEeqnarray}{rCl}
\min_{\{\hat{\mathbf{w}}_{j|k}\}_{k-N-1}^{k-1}}\Phi_k&&=\underbrace{\hat{\mathbf{w}}_{k-N-1|k}^T\mathbf{P}_{k-N}^{-1}\hat{\mathbf{w}}_{k-N-1|k}}_{\text{A Priori State Cost (Arrival Cost)}}+ \nonumber
\\
&&\underbrace{\sum_{j=k-N}^{k-1}\hat{\mathbf{w}}_{j|k}^T\mathbf{Q}_j^{-1}\hat{\mathbf{w}}_{j|k}}_{\text{State Transition Cost}}+\underbrace{\sum_{j=k-N}^{k}\hat{\mathbf{v}}_{j|k}^T\mathbf{R}_j^{-1}\hat{\mathbf{v}}_{j|k}}_{\text{Measurement Cost}}\nonumber
\end{IEEEeqnarray}
\begin{IEEEeqnarray}{rCl}
s.t.:\hat{\mathbf{w}}_{k-N-1|k}&=&\hat{\mathbf{x}}_{k-N|k}-\hat{\mathbf{x}}_{k-N|k-N-1}\nonumber
\\
\hat{\mathbf{w}}_{j|k}&=&\hat{\mathbf{x}}_{j+1|k}-f_j(\hat{\mathbf{x}}_{j|k})\nonumber
\\
\hat{\mathbf{v}}_{j|k}&=&\hat{\mathbf{y}}_{j}-h_j(\hat{\mathbf{x}}_{j|k}) \label{eqn:mhe}
\end{IEEEeqnarray}
where $\hat{[\;]}_{j|k}$ denotes an estimate of an unknown at time $j$ with measurements up to time $k$. The a priori state cost is also called the arrival cost \cite{rao2003constrained}. It summarizes the contribution of the measurements before the current moving horizon \cite{rao2002constrained}. $N$ denotes the horizon size.

If we extend the horizon size $N$ to incorporate all previous measurements, the problem \eqref{eqn:mhe} becomes the batch least squares (BLS) estimation. In that case, the a priori cost describes our confidence in the prior knowledge of the system state:
\begin{equation*}
    \mathbf{x}_{0}\sim N(\bar{\mathbf{x}}_{0}, \mathbf{P}_{0})
\end{equation*}
where $\bar{\mathbf{x}}_{0}=\hat{\mathbf{x}}_{0|-1}$. In the following sections, we will indicate that BLS, EKF, and MHE can be unified.


\section{The Recursive Solution}
Our recursive solution is based on the \textit {linearization} of the NLTV system described by \eqref{eqn:dynEqn} and \eqref{eqn:measEqn}. Our derivation follows the least squares principle \cite{muske1993receding,ling1999receding}. Assume we have the approximate state estimates:
\begin{equation*}
\tilde{\mathbf{X}}_k=\left[\tilde{\mathbf{x}}_{0|0},\tilde{\mathbf{x}}_{1|1},\tilde{\mathbf{x}}_{2|2},\tilde{\mathbf{x}}_{3|3},\cdots,\tilde{\mathbf{x}}_{k|k}\right].
\end{equation*}
Then, we define the a priori state approximation as:
\begin{equation*} \tilde{\mathbf{X}}_{k+1}^{-}=\left[\bar{\mathbf{x}}_{0},\tilde{\mathbf{x}}_{1|0},\tilde{\mathbf{x}}_{2|1},\tilde{\mathbf{x}}_{3|2},\tilde{\mathbf{x}}_{4|3},\cdots,\tilde{\mathbf{x}}_{k+1|k}\right]
\end{equation*}
The dynamic model \eqref{eqn:dynEqn} and the measurement model \eqref{eqn:measEqn} are linearized at $\tilde{\mathbf{X}}_k$ and $\tilde{\mathbf{X}}_{k+1}^{-}$, respectively. Thus, the NLTV system approximates a linear time-variant (LTV) system:
\begin{IEEEeqnarray}{rcl}
\mathbf{x}_{k+1}&=&\mathbf{A}_k\mathbf{x}_k+\mathbf{u}_k+\mathbf{w}_k \label{eqn:linearizedDyn}
\\
\mathbf{y}_k&=&\mathbf{C}_k\mathbf{x}_k+\mathbf{d}_k+\mathbf{v}_k\label{eqn:linearizedMeas}
\end{IEEEeqnarray}
where
\begin{gather}
\mathbf{A}_k=\left[{\partial{f_k}}/{\partial \mathbf{x}_{k}}\right]_{\mathbf{x}_k=\tilde{\mathbf{x}}_{k|k}}, \quad \mathbf{C}_k=\left[{\partial{h_k}}/{\partial \mathbf{x}_{k}}\right]_{\mathbf{x}_k=\tilde{\mathbf{x}}_{k|k-1}}\nonumber
\\
\mathbf{u}_k=f_k(\tilde{\mathbf{x}}_{k|k})-\mathbf{A}_k\tilde{\mathbf{x}}_{k|k}\nonumber
\\
\mathbf{d}_k=h_k(\tilde{\mathbf{x}}_{k|k-1})-\mathbf{C}_k\tilde{\mathbf{x}}_{k|k-1}.\nonumber
\end{gather}
The constraints on system noise in \eqref{eqn:mhe} are linearized correspondingly:
\begin{IEEEeqnarray}{rcl}
\hat{\mathbf{w}}_{j|k}&=&\hat{\mathbf{x}}_{j+1|k}-\mathbf{A}_j\hat{\mathbf{x}}_{j|k}-\mathbf{u}_j\label{eqn:U_noise}
\\
\hat{\mathbf{v}}_{j|k}&=&\mathbf{y}_{j}-\mathbf{C}_j\hat{\mathbf{x}}_{j|k}-\mathbf{d}_j.\label{eqn:V_noise}
\end{IEEEeqnarray}


\begin{lemma}\label{lemma1} Given the prior state estimate $\hat{\mathbf{x}}_{k-N|k-N-1}$ and the state noise estimates given measurements up to time $k$, i.e., $\hat{\mathbf{\Omega}}_N^{k-N}=[\hat{\mathbf{w}}_{k-N-1|k}^T, \hat{\mathbf{w}}_{k-N|k}^T,\cdots,\hat{\mathbf{w}}_{k-1|k}^T]^T$, the state estimate at time $j$ $(k-N \leq j \leq k, k \geq N)$ is:
\begin{IEEEeqnarray}{rcl}
\hat{\mathbf{x}}_{j|k}&=&\mathbf{A}_{j-1}^{j-(k-N)}\hat{\mathbf{x}}_{k-N|k-N-1}\nonumber
\\
&&+\sum_{i=0}^{j-(k-N)-1}\mathbf{A}_{j-1}^{j-(k-N)-i-1}\mathbf{u}_{k-N+i}\nonumber
\\
&&+\sum_{i=0}^{j-(k-N)}\mathbf{A}_{j-1}^{j-(k-N)-i}\hat{\mathbf{w}}_{k-N-1+i|k}\label{eqn:x_expression}
\end{IEEEeqnarray}
where,
\begin{gather*}
\mathbf{A}_{k}^0=\mathbf{I},\quad \mathbf{A}_{-1}^0=\mathbf{I},\quad \mathbf{A}_{k}^N=\prod_{i=0}^{N-1} \mathbf{A}_{k-i},\quad \hat{\mathbf{x}}_{0|-1}=\bar{\mathbf{x}}_0, 
\end{gather*}
\end{lemma}
where $\mathbf{I}$ denotes the identity matrix. Lemma 1 can be proved by induction. 


Substituting \eqref{eqn:U_noise}, \eqref{eqn:V_noise}, and \eqref{eqn:x_expression} into the minimization problem \eqref{eqn:mhe} yields:
\begin{IEEEeqnarray}{rcl}\label{eqn:lsMhe}   &&\min_{\hat{\mathbf{\Omega}}_N^{k-N}}\Phi_k=\hat{\mathbf{\Omega}}_N^{{k-N}^T}\tilde{\mathbf{Q}}_N^{k-N}\hat{\mathbf{\Omega}}_N^{k-N}+
\\
&&\left(\mathbf{B}_N^{k-N}-\mathbf{D}_N^{k-N}\hat{\mathbf{\Omega}}_N^{k-N}\right)^T\mathbf{R}_N^{k-N}\left(\mathbf{B}_N^{k-N}-\mathbf{D}_N^{k-N}\hat{\mathbf{\Omega}}_N^{k-N}\right)\nonumber
\end{IEEEeqnarray}
where the giant covariance matrices describing the system statistics in the current horizon are
\begin{gather}
    \tilde{\mathbf{Q}}_N^{k-N}=diag\{\mathbf{P}_{k-N}^{-1},\mathbf{Q}_{k-N}^{-1},\cdots,\mathbf{Q}_{k-1}^{-1}\}\label{eqn:Q}
    \\
    {\mathbf{R}}_N^{k-N}=diag\{\mathbf{R}_{k-N}^{-1},\mathbf{R}_{k-N+1}^{-1},\cdots,\mathbf{R}_{k}^{-1}\}.\label{eqn:R}
\end{gather}
The giant vector involving all measurements in the horizon is defined as:
\begin{equation}
    \mathbf{B}_N^{k-N}=\left[\mathbf{b}_{k-N}^T,\mathbf{b}_{k-N+1}^T,\cdots,\mathbf{b}_{k}^T\right]^T\label{eqn:B}
\end{equation}
where
\begin{gather}
    \mathbf{b}_j=\mathbf{y}_j-\mathbf{C}_j\mathbf{A}_{j-1}^{j-(k-N)}\hat{\mathbf{x}}_{k-N|k-N-1}-\nonumber
\\
\mathbf{C}_j\sum_{i=0}^{j-(k-N)-1}\mathbf{A}_{j-1}^{j-(k-N)-i-1}\mathbf{u}_{k-N+i}-\mathbf{d}_j.\label{eqn:b}
\end{gather}
The relationship between the measurements and states in the horizon is represented by:
\begin{gather}
\mathbf{D}_N^{k-N}=\label{eqn:D}
\\
\begin{bmatrix}
&\mathbf{C}_{k-N}&0&\cdots&0
\\
&\mathbf{C}_{k-N+1}\mathbf{A}_{k-N}&\mathbf{C}_{k-N+1}&\cdots&0
\\
&\mathbf{C}_{k-N+2}\mathbf{A}_{k-N+1}^2&\mathbf{C}_{k-N+2}\mathbf{A}_{k-N+1}&\cdots&0
\\
&\vdots&\vdots&\vdots&\vdots
\\
&\mathbf{C}_{k-1}\mathbf{A}_{k-2}^{N-1}&\mathbf{C}_{k-1}\mathbf{A}_{k-2}^{N-2}&\cdots&0
\\
&\mathbf{C}_{k}\mathbf{A}_{k-1}^{N}&\mathbf{C}_{k}\mathbf{A}_{k-1}^{N-1}&\cdots&\mathbf{C}_{k}\nonumber
\end{bmatrix}
\end{gather}
The least squares problem described by \eqref{eqn:lsMhe} can be solved as:
\begin{equation}\label{eqn:omega}
\hat{\mathbf{\Omega}}_N^{k-N}=\left(\mathbf{E}_N^{k-N}+\tilde{\mathbf{Q}}_N^{k-N}\right)^{-1}\mathbf{G}_N^{k-N}\mathbf{B}_N^{k-N}
\end{equation}
where
\begin{gather}
\mathbf{E}_N^{k-N}=\mathbf{D}_N^{{k-N}^T}\mathbf{R}_N^{k-N}\mathbf{D}_N^{k-N}\label{eqn:E}
\\
\mathbf{G}_N^{k-N}=\mathbf{D}_N^{{k-N}^T}\mathbf{R}_N^{k-N}.\label{eqn:G}
\end{gather}
Substituting \eqref{eqn:linearizedDyn} and \eqref{eqn:omega} into \eqref{eqn:x_expression} yields the recursive state estimate:
\begin{IEEEeqnarray}{rcl}\label{eqn:solution}
\hat{\mathbf{x}}_{k+1|k}&=&\mathbf{A}_{k}^{N+1}\hat{\mathbf{x}}_{k-N|k-N-1}+\sum_{i=0}^{N}\mathbf{A}_{k}^{N-i}\mathbf{u}_{k-N+i}\nonumber
\\
&&+\mathbf{L}_N^{k-N}\mathbf{B}_N^{k-N}
\end{IEEEeqnarray}
where
\begin{gather}
    \mathbf{L}_N^{k-N}=\tilde{\mathbf{A}}_N^k\left(\mathbf{E}_N^{k-N}+\tilde{\mathbf{Q}}_N^{k-N}\right)^{-1}\mathbf{G}_N^{k-N}\label{eqn:L}
    \\
\tilde{\mathbf{A}}_{N}^{k}=\left[\mathbf{A}_{k}^{N+1},\mathbf{A}_{k}^{N},\cdots,\mathbf{A}_{k}\right].\label{eqn:tildeA} 
\end{gather}



\begin{algorithm}[!t]
\caption{Recursive MHE at time $k$}\label{alg:1}
\KwIn{$\{\mathbf{y}_{k-N},\cdots,\mathbf{y}_k\}$, 
\\
$\quad\quad \{\mathbf{Q}_{k-N},\cdots,\mathbf{Q}_{k-1}\},\{\mathbf{R}_{k-N},\cdots,\mathbf{R}_k\}$,
\\
$\quad\quad \{\hat{\mathbf{x}}_{k-N|k-N-1}^{(N)},\cdots,\hat{\mathbf{x}}_{k|k-1}^{(N)}\}, \{\mathbf{P}_{k-N},\cdots,\mathbf{P}_{k}\}$}

\KwOut{$\hat{\mathbf{x}}_{k+1|k}^{(N)}$}

Linearize the system \eqref{eqn:dynEqn}, \eqref{eqn:measEqn}\;

Prepare system statistics using \eqref{eqn:Q} and \eqref{eqn:R}\;

Prepare measurements using \eqref{eqn:B} and \eqref{eqn:b}\;

Assemble matrices using \eqref{eqn:D}, \eqref{eqn:E}, \eqref{eqn:G}, \eqref{eqn:tildeA}, \eqref{eqn:L}\;

Compute $\hat{\mathbf{x}}_{k+1|k}^{(N)}$ using \eqref{eqn:solution}, and update to $\mathbf{P}_{k+1}$\; 

\Horizon{$\{\mathbf{y}_{k-N+1},\cdots,\mathbf{y}_{k+1}\}$
\\$\quad\{\mathbf{Q}_{k-N+1},\cdots,\mathbf{Q}_{k}\},\{\mathbf{R}_{k-N+1},\cdots,\mathbf{R}_{k+1}\}$,
\\$\{\hat{\mathbf{x}}_{k-N+1|k-N}^{(N)},\cdots,\hat{\mathbf{x}}_{k+1|k}^{(N)}\}, \{\mathbf{P}_{k-N+1},\cdots,\mathbf{P}_{k+1}\}$}
\end{algorithm}

\section{UNIFICATION}
In this section, we unify the classic EKF and popular FGO into the MHE framework. \textit {To the best of our knowledge, our work is the first to provide the conditions under which EKF and MHE achieve equivalence {\bfseries for any horizon size in NLTV systems}.} Additionally, we highlight the connection between FGO and MHE. 

\begin{figure}[!t]
      \centering
      \includegraphics[width=\linewidth]{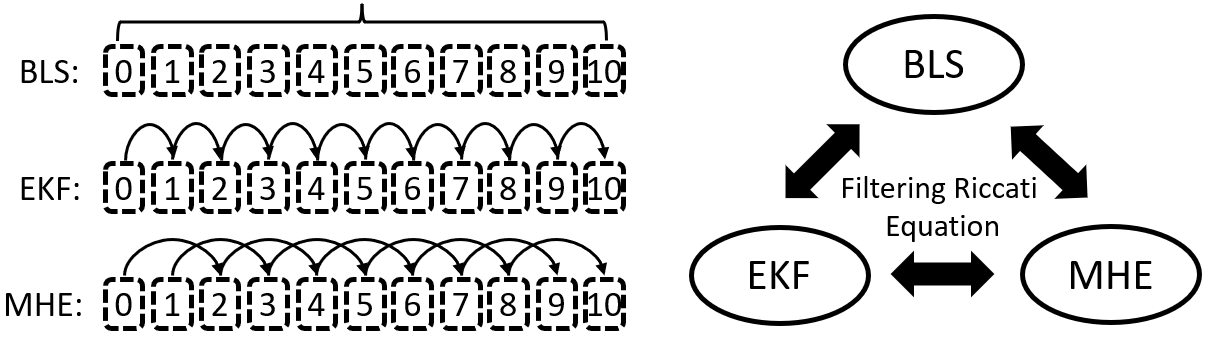}     
      \caption{Threading BLS, EKF, and MHE with the filtering Riccati equation}
      \label{fig:ekf_bls_mhe}
\end{figure}

\subsection{Equivalence between MHE and EKF}
To derive and prove the conditions for the equivalence between EKF and MHE, we followed the idea in \cite{muske1993receding} by extending it from LTI systems to NLTV ones.

\begin{lemma}\label{lemma2}
The filtering algebraic Riccati matrix $\mathbf{P}_k$, which is the solution to
\begin{IEEEeqnarray}{rcl}\label{eqn:riccati}
&&\mathbf{P}_{k}=\mathbf{Q}_{k-1}+\mathbf{A}_{k-1}[\mathbf{P}_{k-1}-
\\
&&\mathbf{P}_{k-1}\mathbf{C}_{k-1}^T\left(\mathbf{C}_{k-1}\mathbf{P}_{k-1}\mathbf{C}_{k-1}^T+\mathbf{R}_{k-1}\right)^{-1}\mathbf{C}_{k-1}\mathbf{P}_{k-1}]\mathbf{A}_{k-1}^T\nonumber
\end{IEEEeqnarray},
can be computed as follows for all $N \geq 0$ and $k>N$:
\begin{equation}\label{eqn:lemma2_right_side}
\mathbf{P}_k=\tilde{\mathbf{A}}_N^{k-1}\left(\mathbf{E}_N^{k-N-1}+\tilde{\mathbf{Q}}_N^{k-N-1}\right)^{-1}\tilde{\mathbf{A}}_N^{{k-1}^T}+\mathbf{Q}_{k-1}.
\end{equation}
\end{lemma}
The proof is given in Appendix \ref{sec:lemma2}.

\begin{corollary}\label{corollary2.1}
From Lemma 2, we can derive the following identity
\begin{equation*}
\left(\mathbf{P}_k^{-1}+\mathbf{C}_k^T\mathbf{R}_k^{-1}\mathbf{C}_k\right)^{-1}=\tilde{\mathbf{E}}_0^{{k}^{-1}}+\mathbf{\Lambda}_N^{k-N-1}
\end{equation*}
where
\begin{gather*}
\tilde{\mathbf{E}}_0^{{k}}=\mathbf{C}_k^T\mathbf{R}_k^{-1}\mathbf{C}_k+\mathbf{Q}_{k-1}^{-1}
\\
\mathbf{\Lambda}_N^{k-N-1}=\tilde{\mathbf{E}}_0^{{k}^{-1}}\mathbf{Q}_{k-1}^{-1}\tilde{\mathbf{A}}_{N}^{k-1}\mathbf{W}_{N}^{{k-N-1}^{-1}}\tilde{\mathbf{A}}_{N}^{{k-1}^T}\mathbf{Q}_{k-1}^{-1}\tilde{\mathbf{E}}_0^{{k}^{-1}}
\\
\mathbf{W}_{N}^{k-N-1}=\mathbf{E}_{N}^{k-N-1}+\tilde{Q}_{N}^{k-N-1}+\tilde{\mathbf{A}}_N^{{k-1}^T}\mathbf{\Theta}_{k-1}\tilde{\mathbf{A}}_N^{{k-1}}
\\
\mathbf{\Theta}_{k-1}=\mathbf{Q}_{k-1}^{-1}-\mathbf{Q}_{k-1}^{-1}\tilde{\mathbf{E}}_0^{{k}^{-1}}\mathbf{Q}_{k-1}^{-1}
\end{gather*}
\end{corollary}

\begin{lemma}\label{lemma3}
For $k\geq N$ and $N > 0$, $\mathbf{L}_N^{k-N}$ can be computed as follows: 
\begin{IEEEeqnarray*}{rcl}
&&\mathbf{L}_{N}^{k-N}=
\\
&&[\left(\mathbf{A}_k-\mathbf{A}_k\mathbf{K_k}\mathbf{C}_k\right)\tilde{\mathbf{A}}_{N-1}^{k-1}\left(\mathbf{E}_{N-1}^{k-N}+\tilde{\mathbf{Q}}_{N-1}^{k-N}\right)^{-1}\mathbf{G}_{N-1}^{k-N},
\\
&&\mathbf{A}_k\mathbf{K}_k].
\end{IEEEeqnarray*}
where 
\begin{equation}\label{eqn:kalmanGain}
    \mathbf{K}_k=(\mathbf{P}_k^{-1}+\mathbf{C}_k^{T}\mathbf{R}_k^{-1}\mathbf{C}_k)^{-1}\mathbf{C}_k^{T}\mathbf{R}_k^{-1}. 
\end{equation}
\end{lemma}
The proof is provided in \ref{sec:lemma3}.

We denote the state estimates of EKF and MHE as $\hat{\mathbf{x}}_{k+1|k}^{(EKF)}$ and $\hat{\mathbf{x}}_{k+1|k}^{(N)}$, respectively. Considering Lemma \ref{lemma2}, Lemma \ref{lemma3}, and Corollary \ref{corollary2.1}, we give the following theorem.

\begin{theorem} \label{the:MheEkf}
If an NLTV system described by \eqref{eqn:dynEqn} and \eqref{eqn:measEqn} is linearized around the following two groups of approximate states, respectively:
\begin{IEEEeqnarray}{rCl}
\tilde{\mathbf{x}}_{k|k}&=&\tilde{\mathbf{x}}_{k|k-1}+\mathbf{K}_k\left(\mathbf{y}_k-h_k(\tilde{\mathbf{x}}_{k|k-1})\right) \label{eqn:taylorDyn}
\\
\tilde{\mathbf{x}}_{k|k-1}&=&\left\{\begin{array}{ll} 
\bar{\mathbf{x}}_0 & \mbox{if $k=0$};
\\
\hat{\mathbf{x}}_{k|k-1}^{(k-1)} & \mbox{if $0<k\leq N$};
\\
\hat{\mathbf{x}}_{k|k-1}^{(N)} & \mbox{if $k> N$}
\end{array}\right.\label{eqn:taylorMeas}
\end{IEEEeqnarray}
where $\mathbf{P}_k$ is the solution of the discrete filtering Riccati equation \eqref{eqn:riccati}, we will have 
\begin{equation*}
\hat{\mathbf{x}}_{k+1|k}^{(N)}=\hat{\mathbf{x}}_{k+1|k}^{(EKF)}, N \in \mathbb{N}.
\end{equation*}
\end{theorem}
The proof is given in Appendix \ref{sec:the1}.

\begin{remark}
Under the conditions specified in \autoref{the:MheEkf}, an algebraic expression for MHE is analytically derived, formulating a filtering-based MHE and establishing the equivalence to EKF accordingly. Since EKF is a recursive solution to the BLS estimation \cite{muske1993receding,robertson1996moving}, MHE can unify EKF and BLS, as shown by Fig. \ref{fig:ekf_bls_mhe}.

\autoref{the:MheEkf} indicates that the nonlinear measurement equation \eqref{eqn:measEqn} at time $k$ should be linearized using the MHE prediction at that time, based on measurements available up to time $k-1$. Note that the horizon size must be truncated if there are not sufficient measurements to construct a complete horizon. The approximate states used to linearize the state transition equation \eqref{eqn:dynEqn} and the measurement equation \eqref{eqn:measEqn} follow a relationship analogous to the update phase in EKF, as demonstrated in \eqref{eqn:taylorDyn}. In fact, the coefficient matrix $\mathbf{K}_k$ is essentially another form of the Kalman gain matrix, which can be proved using the matrix inversion lemma \cite{higham2002accuracy}. Such recursive MHE is summarized in Algorithm \ref{alg:1}. We display here the case where the time $k$ is larger than the horizon size $N$, while other cases can be easily derived from it. 
\end{remark}

\begin{figure*}[!t]
      \centering
      \includegraphics[width=\linewidth]{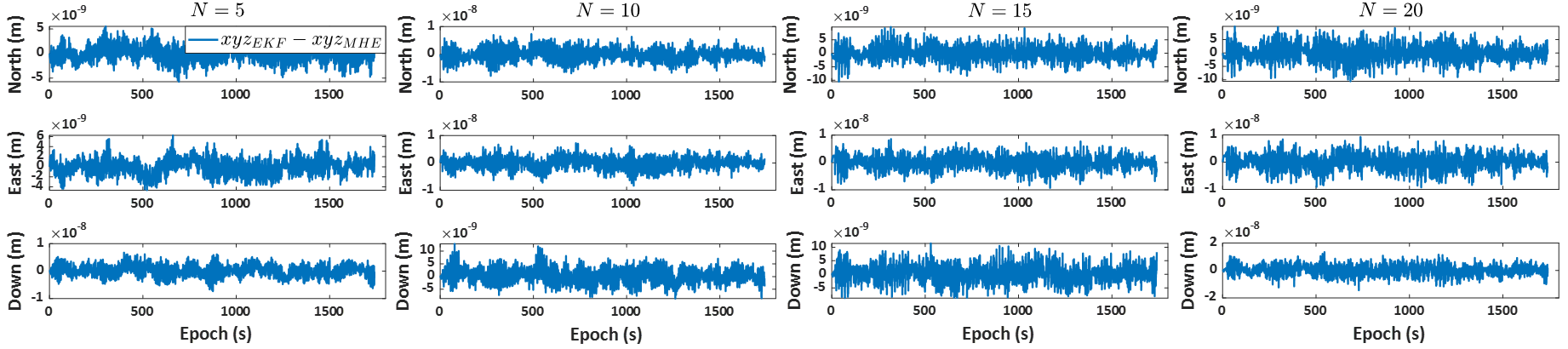}     
      \caption{Difference in state estimates between EKF and MHE under Theorem 1. The nanometer-level residual is the expected floating-point noise. }
      \label{fig:ekf_mhe}
\end{figure*}

Fig. \ref{fig:ekf_mhe} displays the difference in positioning results of EKF and MHE with various horizon sizes, where MHE is implemented according to Theorem \ref{the:MheEkf}. The data is from the GSDC 2021 dataset with an identifier ``2020-05-14-US-MTV-1" \cite{fu2020android}. Fig. \ref{fig:ekf_mhe} illustrates that EKF is equivalent to MHE regardless of the horizon size. The nanometer-level difference between them is due to the expected floating-point noise resulting from their different implementations.

\subsection{Equivalence between MHE and FGO}

\begin{figure}[!t]
      \centering
      \includegraphics[width=0.9\linewidth]{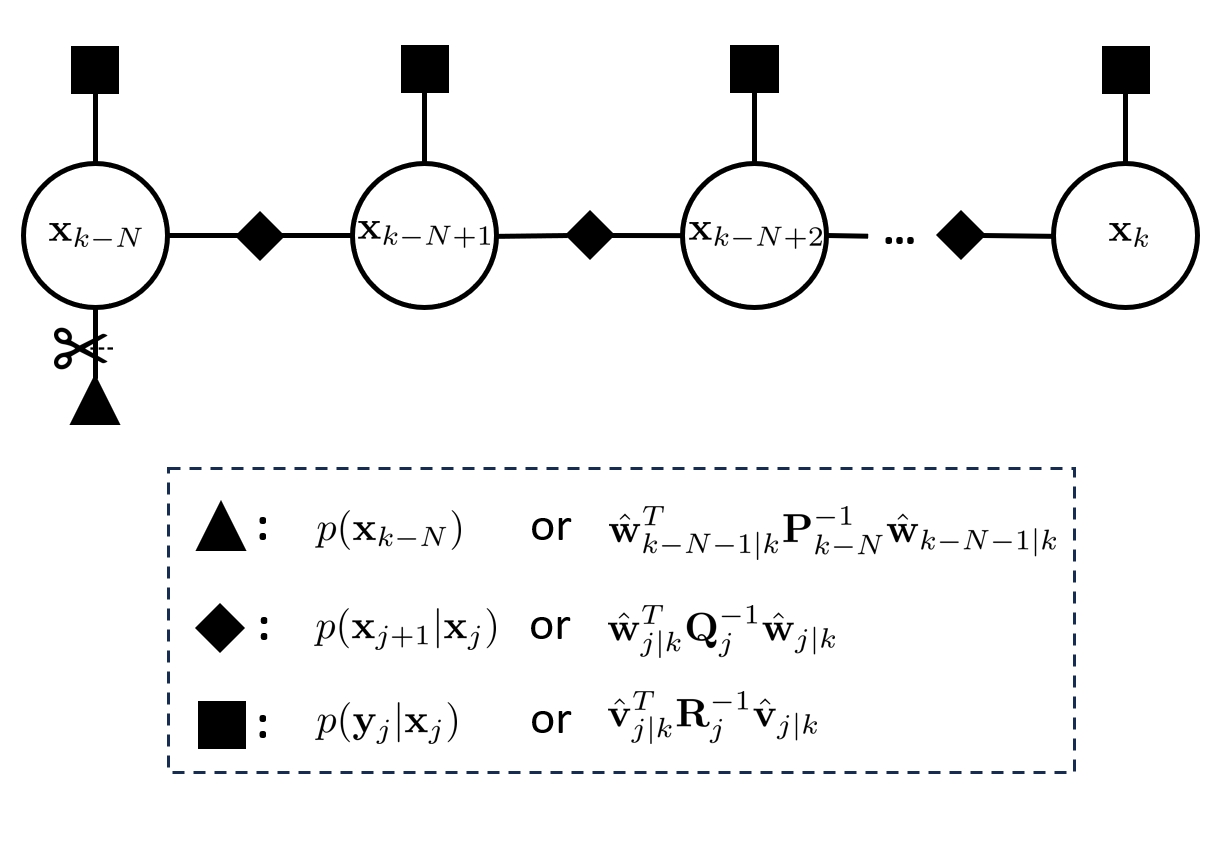}
      
      \caption{Diagram of a basic FGO}
      \label{fig:fgo}
\end{figure}

FGO stems from factorizing a global a posteriori density to perform the maximum a posteriori (MAP) estimation \cite{dellaert2017factor}, which is exactly the same as that done by MHE \cite{muske1993receding}. 


Under the assumption that all the probability density functions in \eqref{eqn: MAP_MHE} are Gaussian distributions defined by \eqref{eqn:noiseAssump}, the MAP problem is equivalent to the least squares problem \eqref{eqn:mhe} \cite{jazwinski1970stochastic}. The corresponding factor graph, denoted by \eqref{eqn: MAP_MHE}, is displayed in Fig. \ref{fig:fgo}. A factor graph has two types of nodes --- the variable nodes, denoted by circles, and the factor nodes, denoted by solid polygons \cite{sunderhauf2012towards,wen2021towards}. The variable nodes represent the unknown states, while the factor nodes describe various relationships among states and known measurements.

{\bfseries Under the Gaussian distribution assumption \eqref{eqn:noiseAssump}}, FGO is equivalent to MHE and can be considered the abstraction and visualization of MHE. However, in GNSS localization, FGO is typically implemented in the absence of the prior state factor, i.e., the triangular node shown in Fig. \ref{fig:fgo} \cite{wen2021factor,zhong2022real,xu2023differentiable}, which is equivalent to omitting the arrival cost in MHE. 


\section{APPLICATION To GNSS LOCALIZATION}
We apply the recursive filtering-based MHE to GNSS localization described by the following NLTV system:
\begin{IEEEeqnarray*}{rcl}
\mathbf{x}_{k+1}&=&\mathbf{A}\mathbf{x}_k+\mathbf{w}_k
\\
\mathbf{y}_k&=&h_k(\mathbf{x}_k)+\mathbf{v}_k
\end{IEEEeqnarray*}
where we try to estimate the user positions $\mathbf{p}_k=[x_k,y_k,z_k]^T$, velocities $\mathbf{v}_k=[v_{x_k},v_{y_k},v_{z_k}]^T$, and time information $[\delta t_{u_k},\delta f_{u_k}]^T$ using ranging measurements $\mathbf{y}_k$. The system states are arranged as follows.
\begin{equation*}
\mathbf{x}_k=\left[x_k, v_{x_k},y_k, v_{y_k},z_k, v_{z_k},\delta t_{u_k},\delta f_{u_k}\right]^T
\end{equation*}

\begin{figure*}[!t]
      \centering
\subfloat[]{\includegraphics[width=0.32\linewidth]{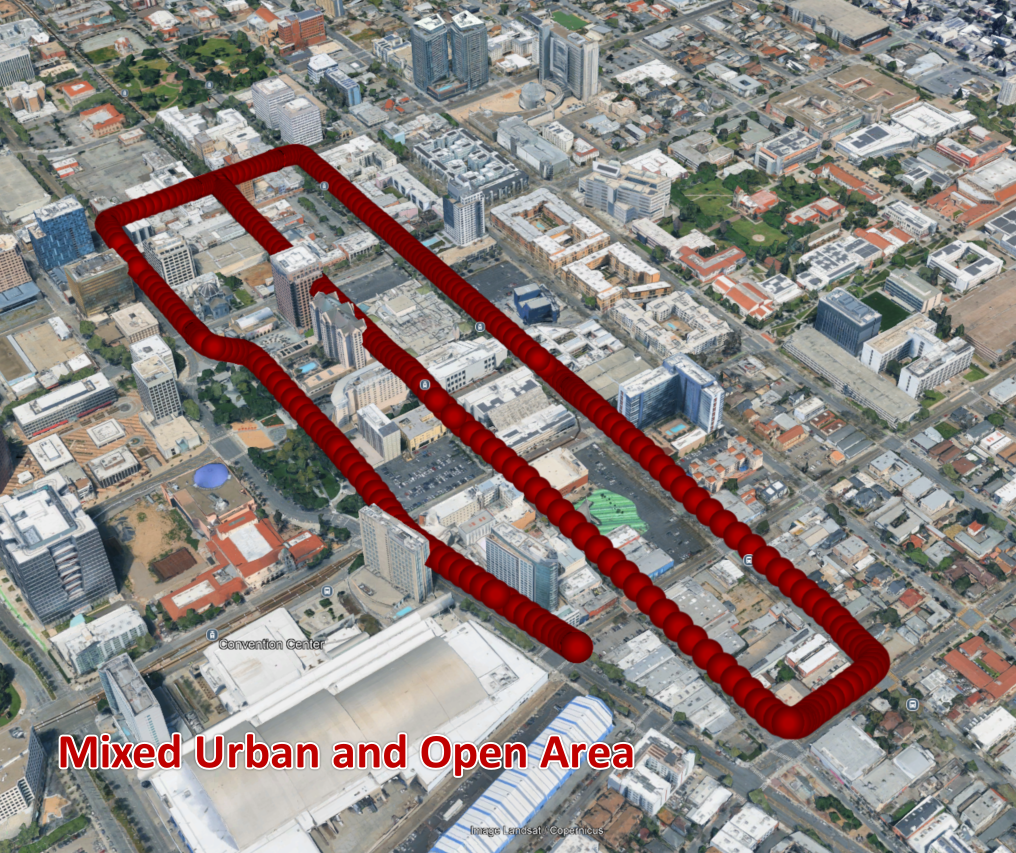}\label{fig:eva_sjc}
}
\hfill
\subfloat[]{\includegraphics[width=0.32\linewidth]{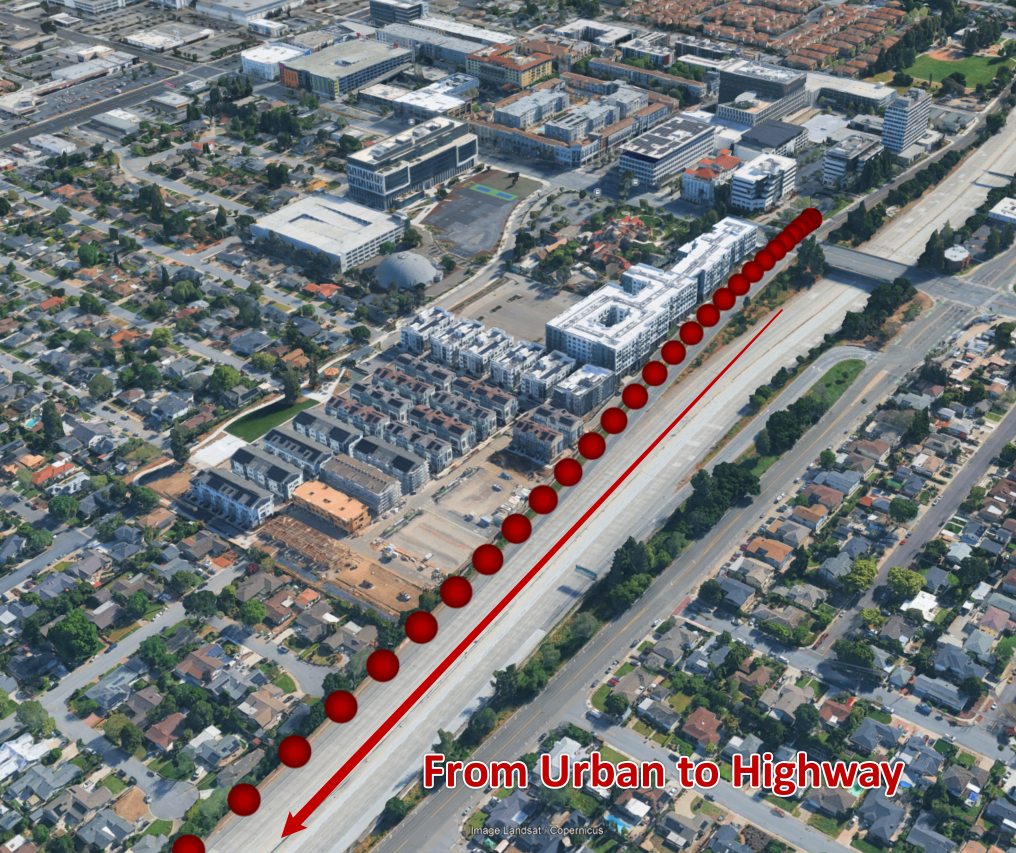}\label{fig:eva_sjc22}
}
\hfill
\subfloat[]{\includegraphics[width=0.32\linewidth]{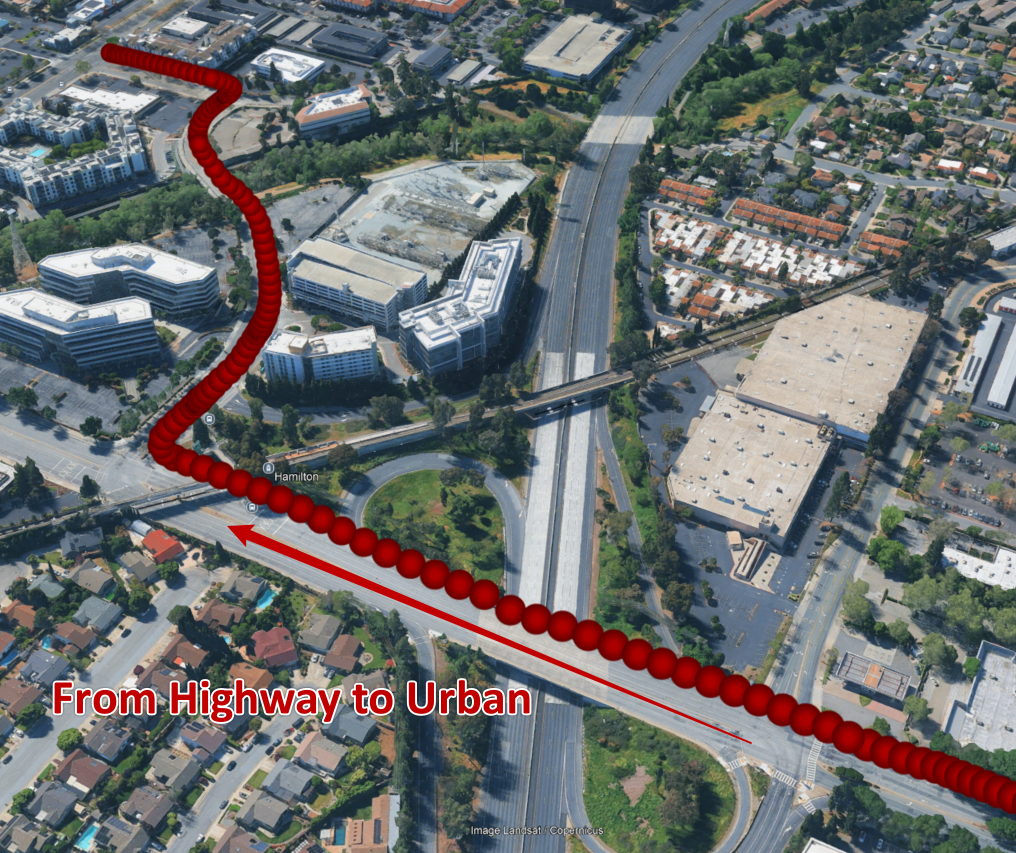}\label{fig:eva_lax}
}
\caption{(a) Scenario 1: a mixed urban and open area; (b) Scenario 2: a transitional area from an urban region to a highway; (c) Scenario 3: a transitional area from a highway to an urban region}
      \label{fig:scene}
\end{figure*}

\begin{figure*}[!t]
      \centering
\subfloat[]{\includegraphics[width=0.32\linewidth]{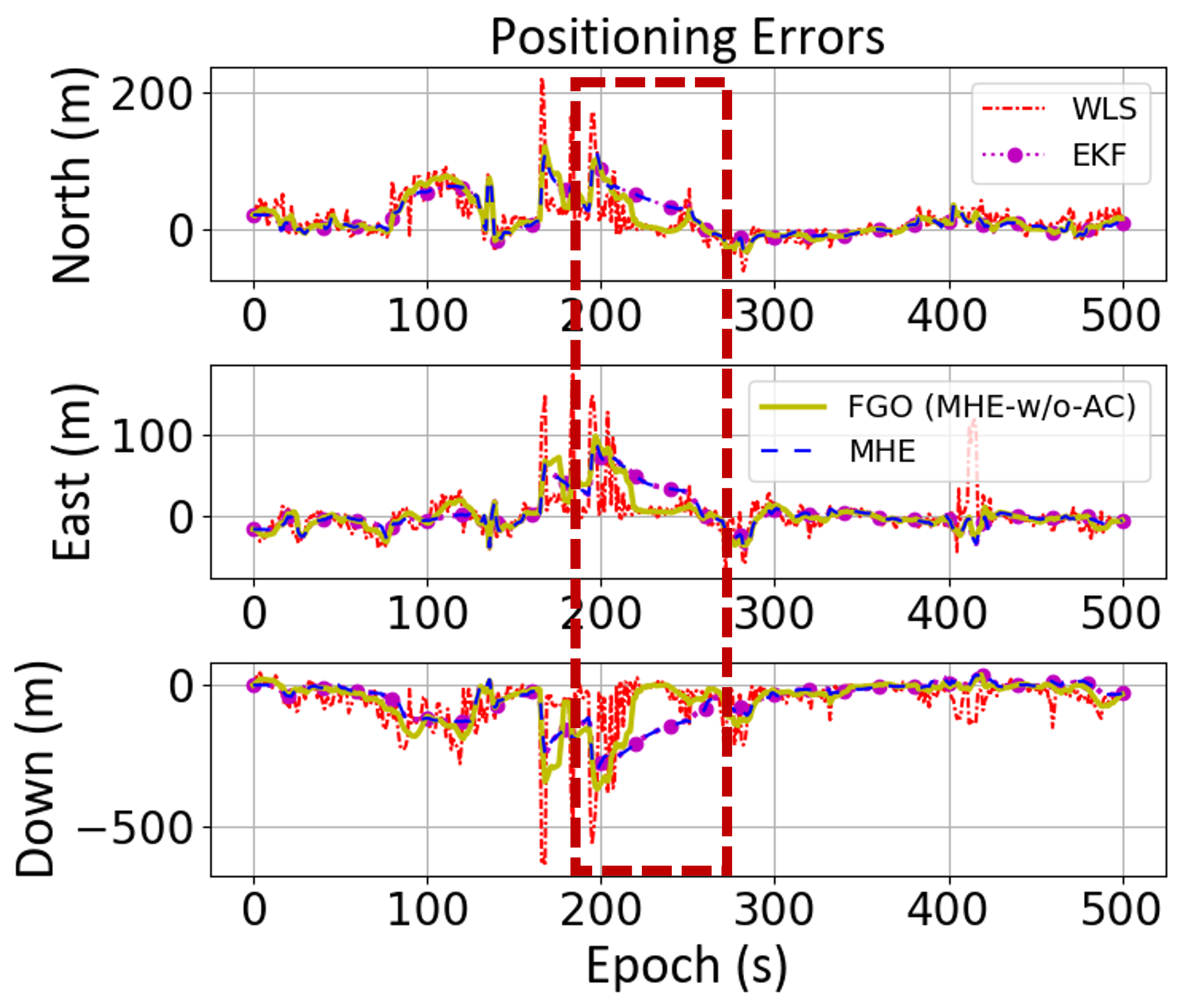}\label{fig: eva_sjc_ned}}
\hfill
\subfloat[]{\includegraphics[width=0.32\linewidth]{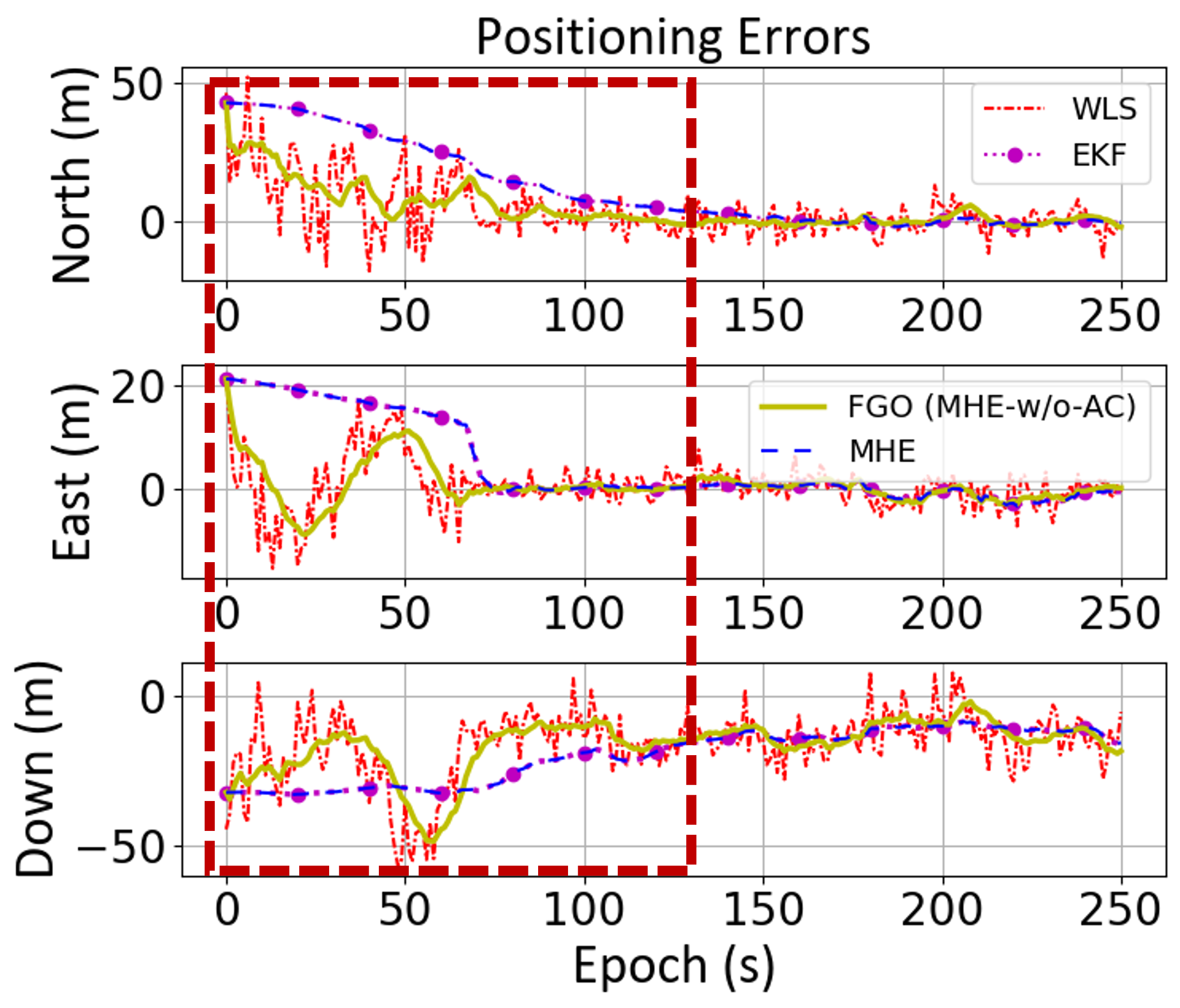}\label{fig:eva_sjc22_ned}}
\hfill
\subfloat[]{\includegraphics[width=0.32\linewidth]{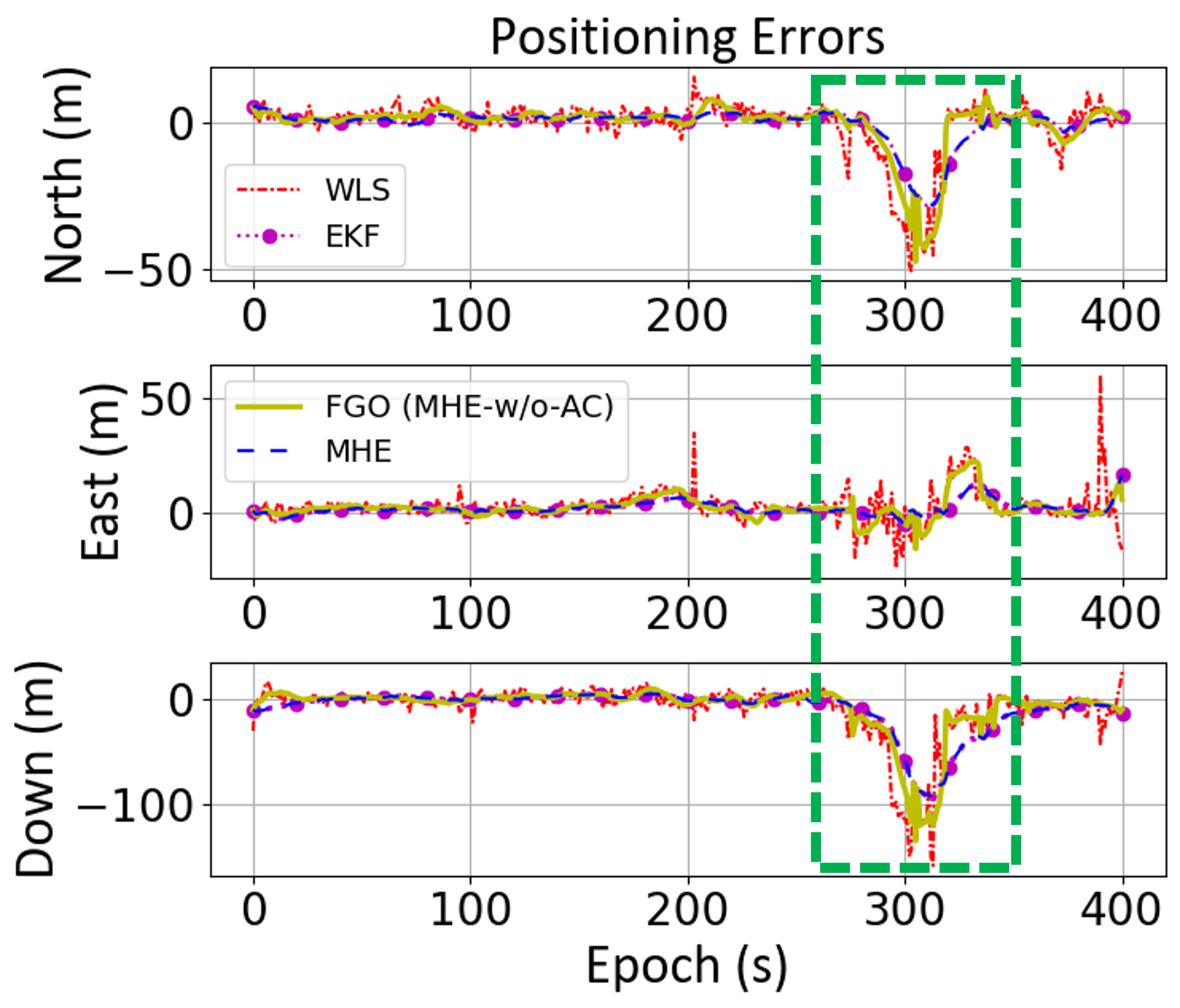}\label{fig: eva_lax_ned}}         
      \caption{Positioning results in Scenario 1 (a), Scenario 2 (b), and Scenario 3 (c)}
      \label{fig:eva}
\end{figure*}

In this paper, we consider the GNSS localization using low-dynamic mobile devices. Therefore, the system dynamics is linearly simplified as
\begin{equation*}
\mathbf{A}=diag\{\mathbf{A}_0,\mathbf{A}_0,\mathbf{A}_0,\mathbf{A}_0\}, \quad \mathbf{A}_0=\begin{bmatrix}
    1&T_{s}
    \\
    0&1
\end{bmatrix}
\end{equation*}
where $T_s$ is the sampling interval of the system. The system measurements are arranged as:
\begin{equation*}
\mathbf{y}_k=\left[\rho_k^{(1)},\dot{\rho}_k^{(1)},\rho_k^{(2)},\dot{\rho}_k^{(2)},\cdots,\rho_k^{(M)},\dot{\rho}_k^{(M)}\right]^T
\end{equation*}
where $M$ is the number of visible GNSS satellites, $\rho_k^{(n)}$ and $\dot{\rho}_k^{(n)}$ denote the pseudorange and pseudorange rate of the $n^{th}$ satellite, respectively. The satellite clock offsets and atmospheric delays in ranging measurements are modeled and eliminated \cite{weng2023localization}. Their nonlinear relationship to the system states, represented by $h_k$, is specified as:
\begin{IEEEeqnarray}{rcl}
\rho_k^{(n)}&=&||\mathbf{p}_k-\mathbf{p}_k^{(n)}||+\delta t_{u_k}\label{eqn:pr}
\\
\dot{\rho}_k^{(n)}&=&(\mathbf{v}_k-\mathbf{v}_k^{(n)})\cdot\mathbf{g}_k^{(n)}+\delta f_{u_k}\label{eqn:prr}
\end{IEEEeqnarray}
where $\mathbf{p}_k^{(n)}$ and $\mathbf{v}_k^{(n)}$ represent the position and velocity of the $n^{th}$ satellite at time $k$, $\mathbf{g}_k^{(n)}$ denotes the unit satellite-user geometry vector: $\mathbf{g}_k^{(n)}=(\mathbf{p}_k-\mathbf{p}_k^{(n)})/{||\mathbf{p}_k-\mathbf{p}_k^{(n)}||}$.
The measurement equations \eqref{eqn:pr} and \eqref{eqn:prr} are linearized according to Theorem \ref{the:MheEkf}, and $\mathbf{g}_k^{(n)}$ is computed using \eqref{eqn:taylorMeas}. We compute the covariance matrix of process noise $\mathbf{Q}_k$ using a two-state model \cite{spilker1996global}. The measurement covariance matrix $\mathbf{R}_k$ is computed using the 1-$\sigma$ uncertainty of ranging measurements \cite{weng2023localization}. The initial state $\bar{\mathbf{x}}_0$ can be initialized using a baseline solution, such as the weighted least squares (WLS) estimate, and the corresponding covariance matrix $\mathbf{P}_0$ is set as a diagonal matrix with 0.05 on the main diagonal.


\section{EXAMPLES}
\subsection{Implementations}
{\bfseries We only evaluate the recursive closed-form solution of MHE described by \eqref{eqn:solution}}. The optimization-based numerical solution of MHE is out of the scope of this paper and will be evaluated in our future work. Similarly, we also implement a recursive MHE without the arrival cost (MHE-w/o-AC) as a filtering-based FGO. A baseline WLS algorithm and a standard EKF are also compared \cite{weng2023localization}. The implementation of MHE follows Theorem \ref{the:MheEkf}. The horizon size of MHE and FGO is set to 10.

\subsection{Datasets and Evaluation Metrics}
To evaluate these estimators, three example trajectories are chosen from the public GSDC 2021, GSDC 2022, and GSDC 2023 datasets, respectively. These data were collected by various phones in San Jose, covering deep urban, suburban, and open highway areas. The ground truth data were collected by Google using the SPAN GNSS Inertial Navigation Systems \cite{fu2020android}. We use them to design three localization scenarios:  (1) mixed downtown urban and open areas (Fig. \ref{fig:eva_sjc}), (2) transitional areas from urban regions to highway (Fig. \ref{fig:eva_sjc22}), and (3) transitional areas from highway to urban regions (Fig. \ref{fig:eva_lax}). The corresponding localization errors in the north-east-down (NED) coordinate system are displayed in Fig. \ref{fig: eva_sjc_ned}, Fig. \ref{fig:eva_sjc22_ned}, and Fig. \ref{fig: eva_lax_ned}. A quantitative comparison of positioning errors is summarized in Table \ref{table_1}, which displays the mean of horizontal errors and the root mean square error (RMSE) in the vertical direction. The horizontal error is computed using Vincenty's formulae, as employed by GSDC.

\begin{table}[t]
\caption{Horizontal (H) and Vertical (V) Positioning Errors (Meters)}
\label{table_1}
\begin{center}
\begin{tabular}{l|l|c|cccc}
\toprule
Datasets & Device & 3D & WLS & EKF & FGO & MHE\\
\hline
&&&&&&\\[-0.8em]
\multirow{2}{*}{GSDC 21} & \multirow{2}{*}{Pixel 4}&H&28.142&{27.335}&\cellcolor{gray!25}\textbf{26.435}&{27.335}
\\
\cline{3-7}
&&&&&&\\[-0.8em]
Scenario 1&&V&62.745&62.330&\cellcolor{gray!25}\textbf{60.440}&62.330\\
\hline
&&&&&&\\[-0.8em]
\multirow{2}{*}{GSDC 22} & \multirow{2}{*}{Pixel 5}&H&8.342&13.765&\cellcolor{gray!25}\textbf{6.003}&13.765
\\
\cline{3-7}
&&&&&&\\[-0.8em]
Scenario 2& &V&16.744&20.036&\cellcolor{gray!25}\textbf{16.491}&20.036\\
\hline
&&&&&&\\[-0.8em]
\multirow{2}{*}{GSDC 23} & \multirow{2}{*}{Pixel 7}&H&8.106&\cellcolor{gray!25}\textbf{4.573}&6.339&\cellcolor{gray!25}\textbf{4.573}
\\
\cline{3-7}
&&&&&&\\[-0.8em]
Scenario 3&Pro &V&13.161&10.785&\cellcolor{gray!25}\textbf{10.427}&10.785\\
\bottomrule
\end{tabular}
\end{center}
\end{table}


\subsection{Localization Accuracy}
{\bfseries According to our findings, no estimation method can always lead others in positioning performance}.

Compared to the baseline WLS algorithm, EKF, FGO, and MHE produce smoother trajectories. This is primarily because all three methods incorporate system dynamics and leverage smoother ranging rate measurements in the estimation process. Additionally, since EKF and MHE are full-information estimators, their trajectories tend to be smoother than those produced by FGO, as shown in Fig. \ref{fig: eva_sjc_ned}, \ref{fig:eva_sjc22_ned}, and \ref{fig: eva_lax_ned}. However, it is important to note that {\bfseries greater smoothness does not necessarily imply higher accuracy}. For instance, in Scenario 2, Table \ref{table_1} shows that the positioning errors of EKF and MHE are actually larger than those of the baseline WLS. Similarly, their errors also exceed those of FGO in Scenarios 1 and 2. 

Why do EKF and MHE perform worse in Scenarios 1 and 2, but outperform others in Scenario 3? Can we improve FGO's performance by increasing the horizon size in Scenario 3? We explore the two questions below.

\subsubsection{\bfseries Impacts of the Arrival Cost} The arrival cost encapsulates the system's historical information \cite{rao2002constrained}, enabling the recursive forms of EKF and MHE to function as full-information estimators, as illustrated in Fig. \ref{fig:ekf_bls_mhe}. {\bfseries While historical information can improve state estimation, it may also introduce adverse effects}. 


In Scenario 3, the user moved from an open area—--where accurate positioning was achieved (before approximately the $250^{th}$ epoch in Fig. \ref{fig: eva_lax_ned})--—into an urban environment. In this region, pervasive multipath and NLOS effects led to significant positioning errors. However, the influence of favorable historical information introduced by the arrival cost mitigated the immediate impact of these errors, reducing the extent of positioning divergence. This effect is highlighted by the green dashed box in Fig. \ref{fig: eva_lax_ned}.


Conversely, when the user transitions from a challenging urban environment to an open area, erroneous historical information can impede the convergence of current state estimates. In such cases, FGO demonstrates greater adaptability by discarding the arrival cost, allowing it to respond more effectively to changes in the environment. This behavior is illustrated by the red dashed boxes in Figs. \ref{fig: eva_sjc_ned} and \ref{fig:eva_sjc22_ned}.

\begin{figure}[!t]
      \centering
\subfloat[]{\includegraphics[width=0.33\linewidth]{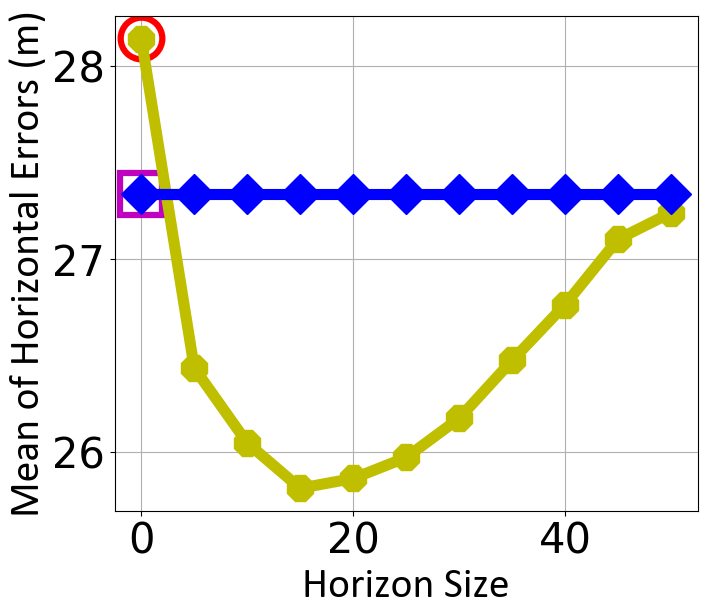}\label{fig:N_test_sjc}
}
\subfloat[]{\includegraphics[width=0.33\linewidth]{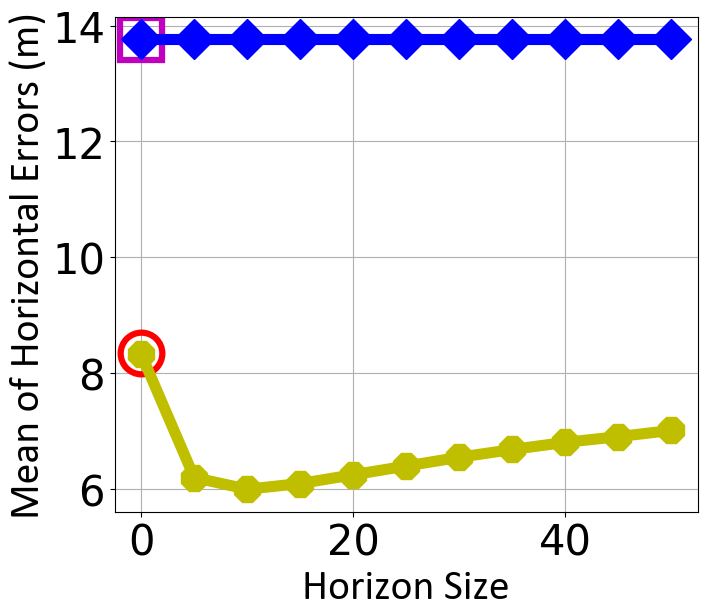}\label{fig: N_test_sjc22}}
\subfloat[]{\includegraphics[width=0.32\linewidth]{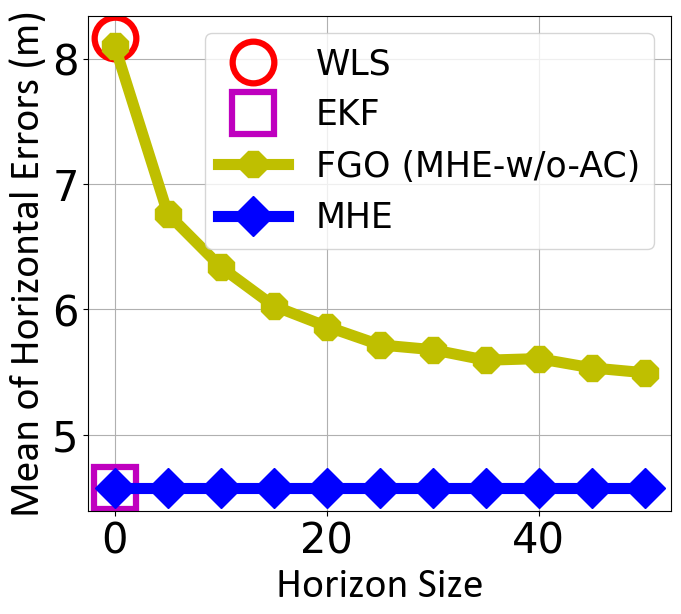}\label{fig:N_test_lax}
}
\caption{Horizontal positioning errors across different horizon sizes in (a) Scenario 1, (b) Scenario 2, and (c) Scenario 3.}
      \label{fig:n_test}
\end{figure}

\subsubsection{\bfseries Impacts of the Horizon Size}
Selecting an appropriate horizon size $N$ is critical for FGO (MHE-w/o-AC), which relies on a fixed-size window of historical data \cite{wen2021factor}. Fig. \ref{fig:n_test} summarizes the horizontal positioning errors of MHE and FGO across varying horizon sizes in the three scenarios.


Since MHE is implemented based on Theorem~\ref{the:MheEkf}, its positioning performance remains consistent across different horizon sizes. Its inferior performance in Scenarios 1 and 2, and superior performance in Scenario 3, are consistent with the earlier analysis regarding the impact of the arrival cost.


Leveraging a small horizon degrades the localization performance of FGO \cite{wen2021factor}, as it becomes more sensitive to noise and outliers in such cases \cite{rawlings2008state}. Increasing the horizon size allows FGO to incorporate more historical data, leading to more robust state estimates. This trend is evident in Fig. \ref{fig:n_test}, where the positioning error initially decreases in Fig.~\ref{fig:N_test_sjc} and Fig.~\ref{fig: N_test_sjc22}, and across the entire range in Fig.~\ref{fig:N_test_lax}.

As the horizon size continues to grow, FGO approaches the behavior of a full-information estimator, causing its positioning performance to converge with that of EKF and MHE. This convergence explains the rebound in FGO’s positioning errors in Fig.~\ref{fig:N_test_sjc} and Fig.~\ref{fig: N_test_sjc22}.

\section{CONCLUSION}
This paper presents a recursive solution of moving horizon estimation for nonlinear time-variant systems. We further establish the conditions under which MHE and EKF are equivalent, regardless of the horizon size. The unification between MHE and FGO is also clarified. All three methods are implemented and compared for GNSS localization.


Our results show that carefully managing the arrival cost and tuning the horizon size can enhance the localization performance of MHE. Given the difficulty of eliminating measurement errors \cite{10506762,10706359}, a context-aware MHE with a learnable arrival cost is a promising research direction \cite{zhang2024gnss}.






\section*{APPENDIX}

\subsection{Proof of Lemma \ref{lemma2}}\label{sec:lemma2}
Lemma \ref{lemma2} can be proved by induction with respect to $N$. Our proof extends the version in \cite{muske1993receding} to the time-variant case. According to the matrix inversion lemma \cite{higham2002accuracy}, the filtering algebraic Riccati matrix \ref{eqn:riccati} is written as:
\begin{equation}\label{eqn:lemma2}
\mathbf{P}_k=\mathbf{A}_{k-1}\left(\mathbf{P}_{k-1}^{-1}+\mathbf{C}_{k-1}^T\mathbf{R}_{k-1}^{-1}\mathbf{C}_{k-1}\right)^{-1}\mathbf{A}_{k-1}^T+\mathbf{Q}_{k-1}.
\end{equation}
Considering $\tilde{\mathbf{A}}_0^{k-1}=\mathbf{A}_{k-1}$, $\mathbf{E}_0^{k-1}=\mathbf{D}_0^{{k-1}^T}\mathbf{R}_0^{k-1}\mathbf{D}_0^{k-1}=\mathbf{C}_{k-1}^T\mathbf{R}_{k-1}^{-1}\mathbf{C}_{k-1}$, and $\tilde{\mathbf{Q}}_0^{k-1}=\mathbf{P}_{k-1}^{-1}$, Lemma 2 is proven for $N=0$. Let $k=N+j$, where $N>0$ and $j>0$. Then, we can assume the following equation holds for $N-1$:
\begin{IEEEeqnarray}{rcl}
    \mathbf{P}_{j+N-1}&=&\mathbf{Q}_{j+N-2}+\label{eqn:induction_P_k_1}
    \\
    &&\tilde{\mathbf{A}}_{N-1}^{j+N-2}\left(\mathbf{E}_{N-1}^{j-1}+\tilde{\mathbf{Q}}_{N-1}^{j-1}\right)^{-1}\tilde{\mathbf{A}}_{N-1}^{{j+N-2}^T}.\nonumber
\end{IEEEeqnarray}
Substituting \eqref{eqn:induction_P_k_1} into \eqref{eqn:lemma2} yields:
\begin{IEEEeqnarray}{rcl}
    \mathbf{P}_{j+N}&=&\mathbf{Q}_{j+N-1}+\label{eqn:lemma2_inter}
    \\
    &&\mathbf{A}_{j+N-1}\left[\tilde{\mathbf{E}}_0^{{j+N-1}^{-1}}+\mathbf{\Lambda}_{N-1}^{j-1}\right]\mathbf{A}_{j+N-1}^T\nonumber
\end{IEEEeqnarray}

On the other hand, considering \eqref{eqn:E} can be computed recursively, i.e.:
\begin{equation}\label{eqn:RecurE}
    \mathbf{E}_N^{k-N}=\begin{bmatrix}
\mathbf{E}_{N-1}^{k-N}+\tilde{\mathbf{A}}_{N-1}^{{k-1}^T}\mathbf{F}_0^{k}\tilde{\mathbf{A}}_{N-1}^{{k-1}}&\tilde{\mathbf{A}}_{N-1}^{{k-1}^T}\mathbf{F}_0^{k}
\\
\mathbf{F}_0^{k}\tilde{\mathbf{A}}_{N-1}^{{k-1}}&\mathbf{F}_0^{k}
\end{bmatrix},
\end{equation}
where $\mathbf{F}_0^{k}=\mathbf{C}_k^T\mathbf{R}_k^{-1}\mathbf{C}_k$, the right side of \eqref{eqn:lemma2_right_side} is written as:
\begin{equation*}
    \tilde{\mathbf{A}}_N^{j+N-1}\begin{bmatrix}
        \boldsymbol{\mathcal{E}}_{N-1}^{j-1}&\mathbf{F}_{N-1}^{j+N-1}
        \\
        \mathbf{F}_{N-1}^{{j+N-1}^T}&\tilde{\mathbf{E}}_0^{j+N-1}
    \end{bmatrix}^{-1}\tilde{\mathbf{A}}_N^{{j+N-1}^T}+\mathbf{Q}_{j+N-1},
\end{equation*}
where 
\begin{gather*}
    \boldsymbol{\mathcal{E}}_{N-1}^{j-1}=\mathbf{E}_{N-1}^{j-1}+\tilde{\mathbf{A}}_{N-1}^{{j+N-2}^T}\mathbf{F}_0^{j+N-1}\tilde{\mathbf{A}}_{N-1}^{j+N-2}+\tilde{\mathbf{Q}}_{N-1}^{j-1}
    \\
    \mathbf{F}_{N-1}^{j+N-1}=\tilde{\mathbf{A}}_{N-1}^{{j+N-2}^T}\mathbf{F}_0^{j+N-1},
\end{gather*}
which can be reformulated exactly the same as the right side of \eqref{eqn:lemma2_inter} according to the formula of blockwise inversion \cite{lu2002inverses}. Therefore, Lemma \ref{lemma2} also holds for $N$, which proves the lemma by induction.


\subsection{Proof of Lemma \ref{lemma3}}\label{sec:lemma3}
 Substituting \eqref{eqn:RecurE} into \eqref{eqn:L} and expanding the expression with the formula for block matrix inversion \cite{lu2002inverses} yield:
\begin{IEEEeqnarray}{rcl}
    \mathbf{L}_N^{k-N}&=&\mathbf{A}_k[\tilde{\mathbf{E}}_0^{k^{-1}}\mathbf{Q}_{k-1}^{-1}\tilde{\mathbf{A}}_{N-1}^{k-1}\mathbf{W}_{N-1}^{k-N}\mathbf{G}_{N-1}^{k-N},\nonumber
    \\
    &&(\mathbf{\Lambda}_{N-1}^{k-N}+\tilde{\mathbf{E}}_0^{k^{-1}})\mathbf{G}_0^k]\label{eqn:lemma3_inter}
\end{IEEEeqnarray}
We separately handle each component of $\mathbf{L}_N^{k-N}$ shown by \eqref{eqn:lemma3_inter}. By performing the matrix inversion lemma \cite{higham2002accuracy} on $\mathbf{W}_{N-1}^{k-N}$ and applying Lemma \ref{lemma2}, we can rewrite the first component as:
\begin{IEEEeqnarray*}{rcl}
    &&\tilde{\mathbf{E}}_0^{k^{-1}}\mathbf{Q}_{k-1}^{-1}\tilde{\mathbf{A}}_{N-1}^{k-1}\mathbf{W}_{N-1}^{k-N}\mathbf{G}_{N-1}^{k-N}=
    \\
    &&\left[\mathbf{I}-\mathbf{K}_k\mathbf{C}_k\right]\tilde{\mathbf{A}}_{N-1}^{k-1}\left(\mathbf{E}_{N-1}^{{k-N}}+\tilde{\mathbf{Q}}_{N-1}^{k-N}\right)\mathbf{G}_{N-1}^{k-N}.
\end{IEEEeqnarray*}
Then, applying Corollary \ref{corollary2.1} to the second component yields:
\begin{equation*}
    (\mathbf{\Lambda}_{N-1}^{k-N}+\tilde{\mathbf{E}}_0^{k^{-1}})\mathbf{G}_0^k=\mathbf{K}_k,
\end{equation*}
which proves Lemma \ref{lemma3}.

\subsection{Proof of Theorem \ref{the:MheEkf}}\label{sec:the1}
The proof is by induction with respect to $N$. For $N=0$ and $k\geq 0$, according to the conditions described by \eqref{eqn:taylorDyn} and \eqref{eqn:taylorMeas}, the equation \eqref{eqn:L} is reformulated as:
\begin{IEEEeqnarray*}{rcl}
   \hat{\mathbf{x}}_{k+1|k}^{(0)}&=&\mathbf{A}_{k}\hat{\mathbf{x}}_{k|k-1}^{(0)}+\mathbf{u}_{k}+\mathbf{L}_0^{k}\mathbf{B}_0^{k}
\\
&=&\mathbf{A}_{k}\left(\hat{\mathbf{x}}_{k|k-1}^{(0)}+\mathbf{K}_k\mathbf{b}_k\right)+\mathbf{u}_k
\\
&=&f_k\left(\hat{\mathbf{x}}_{k|k-1}^{(0)}+\mathbf{K}_k\left(\mathbf{y}_k-h_k(\hat{\mathbf{x}}_{k|k-1}^{(0)})\right)\right)=\hat{\mathbf{x}}_{k+1|k}^{(EKF)} 
\end{IEEEeqnarray*}
Thus, we can assume that Theorem \ref{the:MheEkf} holds for $N=M\geq 0$:
\begin{IEEEeqnarray*}{rcl}
\hat{\mathbf{x}}_{k+1|k}^{(M)}&=&f_k\left(\hat{\mathbf{x}}_{k|k-1}^{(M)}+\mathbf{K}_k\left(\mathbf{y}_k-h_k(\hat{\mathbf{x}}_{k|k-1}^{(M)})\right)\right)
\\
&=&\hat{\mathbf{x}}_{k+1|k}^{(EKF)}
\end{IEEEeqnarray*}

Then, we continue to prove the theorem for $N=M+1$ on a case-by-case basis.

1) When $k=0$, we can obtain the following equality:
\begin{equation*}  \hat{\mathbf{x}}_{1|0}^{(M+1)}=\hat{\mathbf{x}}_{1|0}^{(0)}=\hat{\mathbf{x}}_{1|0}^{(EKF)}
\end{equation*}

2) When $0<k\leq M+1$, considering $\hat{\mathbf{x}}_{k|k-1}^{(M+1)}=\hat{\mathbf{x}}_{k|k-1}^{(M)}=\hat{\mathbf{x}}_{k|k-1}^{(k-1)}$, we can get the following equality by substituting Lemma \ref{lemma3} to \eqref{eqn:x_expression}:
\begin{IEEEeqnarray*}{rcl}
\hat{\mathbf{x}}_{k+1|k}^{(M+1)}&=&\mathbf{A}_{k}^{k+1}\bar{\mathbf{x}}_{0}+\sum_{i=0}^{k}\mathbf{A}_{k}^{k-i}\mathbf{u}_{i}+\mathbf{L}_{k}^{0}\mathbf{B}_{k}^{0}
\\
&=&f_k(\hat{\mathbf{x}}_{k|k-1}^{(M)}+\mathbf{K}_k\left[\mathbf{y}_k-h_k(\hat{\mathbf{x}}_{k|k-1}^{(M)})\right])=\hat{\mathbf{x}}_{k+1|k}^{(EKF)}  
\end{IEEEeqnarray*}

3) When $k>M+1$, we use induction again but with respect to $k$ to prove the equivalence. If $k=M+2$, according to \eqref{eqn:L}, Lemma \ref{lemma3}, conditions \eqref{eqn:taylorDyn}\eqref{eqn:taylorMeas}, and case (2), we have:
\begin{equation*}
\hat{x}_{M+3|M+2}^{(M+1)}=\hat{\mathbf{x}}_{M+3|M+2}^{(EKF)}=\hat{x}_{M+3|M+2}^{(M)}
\end{equation*}


Therefore, we can assume the following equality holds for $k=\eta >M+1$:
\begin{equation}
\hat{x}_{\eta+1|\eta}^{(M+1)}=\hat{\mathbf{x}}_{\eta+1|\eta}^{(EKF)}=\hat{\mathbf{x}}_{\eta+1|\eta}^{(M)}\label{eqn:induction_inter_eta}
\end{equation}



Then, for $k=\eta+1$, according to \eqref{eqn:L}, Lemma \ref{lemma3}, conditions \eqref{eqn:taylorDyn}\eqref{eqn:taylorMeas}, and \eqref{eqn:induction_inter_eta}, we have:
\begin{IEEEeqnarray*}{rcl}
\hat{\mathbf{x}}_{\eta+2|\eta+1}^{(M+1)}&=&\mathbf{A}_{\eta+1}^{M+2}\hat{\mathbf{x}}_{\eta-M|\eta-M-1}^{(M+1)}+
\\
&&\sum_{i=0}^{M+1}\mathbf{A}_{\eta+1}^{M+1-i}\mathbf{u}_{\eta-M+i}+\mathbf{L}_{M+1}^{\eta-M}\mathbf{B}_{M+1}^{\eta-M}
\\
&=&f_{\eta+1}(\hat{\mathbf{x}}_{\eta+1|\eta}^{(M)}+\mathbf{K}_{\eta+1}\left[\mathbf{y}_{\eta+1}-h_{\eta+1}(\hat{\mathbf{x}}_{\eta+1|\eta}^{(M)})\right])
\\
&=&\hat{\mathbf{x}}_{\eta+2|\eta+1}^{(EKF)}.
\end{IEEEeqnarray*}

Therefore, Theorem \ref{the:MheEkf} also holds for $k>M+1$, which proves the theorem for $N=M+1$. The theorem is proved.

\bibliography{main}
\bibliographystyle{IEEEtran}

\end{document}